\documentclass{aastex62}

\shorttitle{Calibration of GRB correlations}
\shortauthors{Wang \& Wang}

\usepackage{graphicx}   
\usepackage{amsmath}    
\usepackage{amssymb}    
\usepackage{cases}
\newcommand{\keV}{\text{ keV}}
\newcommand{\erg}{\text{ erg}}
\newcommand{\km}{\text{ km}}
\newcommand{\mpc}{\text{ Mpc}}
\newcommand{\second}{\text{ s}}
\newcommand{\yr}{\text{ yr}}
\newcommand{\ud}{\mathrm{d}}

\newcommand{\p}{\partial}

\begin{document}
    \title{Calibration of gamma-ray bursts luminosity correlations using gravitational waves as standard sirens}

    \correspondingauthor{F. Y. Wang}
    \email{fayinwang@nju.edu.cn}

    \author{Y. Y. Wang}
    \affil{School of Astronomy and Space Science, Nanjing University, Nanjing 210093, China}

    \author{F. Y. Wang}
    \affil{School of Astronomy and Space Science, Nanjing University, Nanjing 210093, China}
    \affiliation{Key Laboratory of Modern Astronomy and Astrophysics (Nanjing University), Ministry of Education, Nanjing 210093, China}

    \begin{abstract}

Gamma-ray bursts (GRBs) are a potential tool to probe high-redshift
universe. However, the circularity problem enforces people to find
model-independent methods to study the luminosity correlations of
GRBs. Here, we present a new method which uses gravitational waves
as standard sirens to calibrate GRB luminosity correlations. For the
third-generation ground-based GW detectors (i.e., Einstein
Telescope), the redshifts of gravitational wave (GW) events
accompanied electromagnetic counterparts can reach out to $\sim 4$,
which is more distant than type Ia supernovae ($z\lesssim 2$). The
Amati relation and Ghirlanda relation are calibrated using mock GW
catalogue from Einstein Telescope. We find that the $1\sigma$
uncertainty of intercepts and slopes of these correlations can be
constrained to less than 0.2\% and 8\% respectively. Using
calibrated correlations, the evolution of dark energy equation of
state can be tightly measured, which is important for discriminating
dark energy models.

    \end{abstract}

    \keywords{gamma-ray burst: general--gravitational waves: standard sirens}

\section{Introduction}
Gamma-ray burst (GRB) are one of the most energetic phenomena in
our Universe \citep{Kumar15,Wang15}. The high luminosity makes them
detectable out to high redshifts. Therefore, GRBs are promising tool
to probe the high-redshift universe: including the cosmic expansion
and dark energy \citep{Dai04,Liang06,Schaefer07,Wang07}, star
formation rate \citep{Totani97,Bromm02,Wang09,Wang13}, the
reionization epoch \citep{Barkana04,Totani06} and the metal
enrichment history of the Universe \citep{Wang12,Hartoog15}. Among them, the
$\gamma$-ray bursts correlations \citep[for reviews,
see][]{Wang15,Dainotti17,Dainotti18a,Dainotti18b} are most widely studied, which can not
only shed light on the radiation mechanism of GRBs, but also provide
a promising tool to probe the cosmic expansion and dark energy
\citep{Wang15,Dainotti17}. These correlations can be divided into
three categories, such as prompt correlations, afterglow
correlations and prompt-afterglow correlations. The prompt
correlations mainly include Amati correlation \citep{Amati02},
Ghirlanda correlation \citep{Ghirlanda04a}, Liang-Zhang correlation
\citep{Liang05}, Yonetoku correlation \citep{Wei03,Yonetoku04} and
$L_{\text{iso}}-\tau_{\text{lag}}$ correlation \citep{Norris00}.
Afterglow correlations contain only parameters in the afterglow
phase, such as Dainotti correlation ($L_X(T_a)-T^*_{X,a}$)
\citep{Dainotti08}, $L_X(T_a)-T^*_{X,a}$ and $L_O(T_a)-T^*_{O,a}$
correlations \citep{Ghisellini09} and
$L_{O,200\text{s}}-\alpha_{O,>200\text{s}}$ correlation
\citep{Oates12}. Prompt-afterglow correlations connect plateaus and
prompt phases, referring to $E_{\gamma, \text{afterglow}}-E_{X,
\text{prompt}}$ correlation \citep{Liang07}, $L_{X,
\text{afterglow}}-E_{\gamma, \text{prompt}}$ correlation
\citep{Berger07}, $L_X(T_a)-L_{\gamma, \text{iso}}$ correlation
\citep{Dainotti11}, $L_{\text{iso}}-E_{\text{p},z}-\Gamma_0$ correlation \citep{Liang15} and so on.

However, there is a circularity problem when treating GRBs as
relative standard candles. It arises from the derivation of
quantities like luminosity $L_{\text{iso}}$, isotropic energy
$E_{\text{iso}}$ and collimation-corrected energy $E_\gamma$, which
are dependent on luminosity distance $d_L$ in a fiducial cosmology. For instance, the
$d_L$ in a flat $\Lambda$CDM model can be expressed as
\begin{equation}
d_L(z) = \frac{c(1+z)}{H_0}\int_{0}^{z}\frac{\ud z'}{\sqrt{\Omega_{\text{m}}(1+z')^3 + (1-\Omega_m-\Omega_\Lambda)(1+z')^2+\Omega_\Lambda}}.
\label{eqn:1}
\end{equation}
 Therefore, it is
inappropriate to use the model-dependent luminosity correlations to
study cosmology models in turn. Several approaches have been
proposed to overcome the problem \citep{Wang15,Dainotti17}. One
method is to fit the cosmological parameters and luminosity
correlation simultaneously \citep{Ghirlanda04b,Li08}. Another method
is to calibrate the correlations using type Ia supernovae (SNe Ia)
\citep{Liang08} or observational Hubble Data (OHD) \citep{Amati18}. It is based on the
principle that objects of the same redshift should have same luminosity
distance. \cite{Wang08} pointed out that the GRB luminosity
correlations calibrated by SNe Ia are no longer completely
independent of the SNe Ia data points. Consequently, the GRB data
cannot be combined with the SNe Ia dataset directly to constrain
cosmological parameters. What's worse, high redshift SNe Ia can
hardly be found and the furthest SN Ia yet seen is GND12Col with $z
= 2.26^{+0.02}_{-0.10}$ \citep{Rodney15}, while the redshift of GRB
can be up to 9.4 \citep{Cucchiara11}. Moreover, there are many some
systematic uncertainties for SNe Ia, such as dust in the light path \citep{Avgoustidis09,Hu17}, the possible intrinsic evolution \textbf{of}
SN luminosity, magnification by gravitational lensing \citep{Holz98},
peculiar velocity \citep{Hui06}, and so on. These processes
will degrade the usefulness of SNe Ia as standard candles.

Here, we come up with the idea to calibrate GRB luminosity relations
using gravitational waves (GW) standard sirens. The detection of
GW170817 accompanied by electromagnetic counterparts heralds the new
era of gravitational-wave multi-messenger astronomy
\citep{Abbott17}. \cite{Schutz86} first pointed out that the
waveform signals from inspiralling compact binaries can be used to
determine the luminosity distance to the source, serving as a
standard siren. This kind of standard siren is a self-calibrating
distance indicator, which just relying on the modelling of the
two-body problem in general relativity \citep{Sathyaprakash10}. The
of detected BNS and BH-NS merger events can reach up to $z\sim 4$ to
the farthest by Einstein Telescope (ET)
\citep{Abernathy11,Li15,Cai17}, going beyond the redshift limitation
of SNe Ia. For the third generation detectors, such as ground-based
Einstein Telescope (ET) \citep{Abernathy11}, space-based Big Bang
Observer (BBO) \citep{Cutler09}, and Deci-Hertz Interferometer
Gravitational wave Observatory (DECIGO) \citep{Kawamura11}, smaller
distance uncertainty will be achieved than Advanced LIGO and Virgo
\citep{Abbott17}.

The paper is organized as follows. In Section 2, we introduce the
procedure of construction mock GW catalogue. The calibration of GRB
luminosity correlations with GW standard sirens is illustrated in
Section 3. A summary of our result and future outlooks is provided
in the end of the paper.

\section{Construction of GW standard sirens}
\subsection{Redshift distribution}
In order to construct a mock GW catalogue, we need to consider the redshift distribution of the sources, which satisfies the following expression
\begin{equation}
P(z)\propto \frac{4\pi d_C^2(z)R(z)}{H(z)(1+z)},
\end{equation}
where $d_C(z)$ is the comoving distance of the source. The time evolution of the NS-NS merger rate $R(z)$ is given by \citep{Schneider01}
\begin{equation}
    R(z)=
    \begin{cases}
    1+2z, & z\le 1 \\
    \frac{3}{4}(5-z), & 1<z<5 \\
    0, & z\ge 5.
    \end{cases}
\end{equation}
The NS-NS merger rate at redshift $z$ is $\dot{n}(z) =
\dot{n}_0\cdot R(z)$ and the merger rate today is about $\dot{n}_0 = 1.54^{+3.2}_{-1.22}\times 10^{-6}\mpc^{-1}\yr^{-1}$
\citep{Abbott17}.

\subsection{Simulation of luminosity distances}
It is necessary to define the total mass $M_{\text{phys}} = m_1 +
m_2$, symmetric mass ratio $\eta = \frac{m_1 m_2}{M^2}$ and chirp
mass $\mathcal{M}_{c,\text{obs}} = M\eta^{3/5}$ before our analysis,
given binary component masses $m_1$ and $m_2$. The observed chirp
mass is related to physical chirp mass via
$\mathcal{M}_{c,\text{obs}}= (1+z)\mathcal{M}_{c,\text{phys}}$. Similarly, the
observed total mass is $M_{\text{obs}} = (1+z)M_{\text{phys}}$.

\subsubsection{Frequency domain waveform and Fourier Amplitude}
The response of the detector $h(t)$ is a linear combination of two components
\begin{equation}
    h(t) = F_{+}(\theta, \phi, \psi) h_{+}(t) + F_{\times}(\theta, \phi, \psi) h_\times(t),
\end{equation}
where $F_{+}$ and $F_{\times}$ are the antenna pattern functions of
the detector, $\psi$ is the polarization angle, and ($\theta, \phi$) is
the location of the source on the sky. The antenna pattern functions
of ET are
\begin{equation}
\begin{aligned}
F_{+}^{(1)}(\theta, \phi, \psi) &= \frac{{\sqrt 3 }}{2}\bigg[\frac{1}{2}(1 + \cos^2\theta) \cos 2\phi \cos 2\psi  - \cos\theta \sin 2\phi \sin 2\psi \bigg],\\
F_{\times}^{(1)}(\theta, \phi, \psi) &= \frac{{\sqrt 3 }}{2}\bigg[\frac{1}{2}(1 + \cos ^2\theta )\cos 2\phi \sin 2\psi  + \cos\theta \sin 2\phi \cos 2\psi \bigg].
\end{aligned}
\end{equation}
The other pattern functions are $F_{+,\times}^{(2)}(\theta, \phi,
\psi) = F_{+,\times}^{(1)}(\theta, \phi+2\pi/3, \psi)$ and
$F_{+,\times}^{(3)}(\theta, \phi, \psi) = F_{+,\times}^{(1)}(\theta,
\phi+4\pi/3, \psi)$ respectively. The Fourier transform of time
domain waveform $h(t)$ is given by
\begin{equation}
    \mathcal{H}(f) = \mathcal{A} f^{-7/6}\exp(\text{i}(2\pi f t_0 - \pi/4 + 2\psi(f/2)-\phi_{(2,0)})),
\end{equation}
where
\begin{equation}
    \mathcal{A} = \frac{1}{d_L}\sqrt{F^2_+ (1+\cos^2\iota)^2 + 4 F^2_{\times}\cos^2\iota}\sqrt{\frac{5\pi}{96}}\pi^{-7/6} \mathcal{M}_{c,\text{obs}}^{5/6}.
\end{equation}
is the Fourier amplitude. The post-Newtonian formalism of GW
waveform phase up to 3.5 PN is used \citep{Blanchet02} and the
expressions of functions $\psi$ and $\phi_{(2,0)}$ can be found in
\cite{Arun05} and \cite{Zhao11}.

The component masses of binary neutron stars are randomly sampled in
[1,2] $M_\odot$, while for neutron star-black hole systems, the
component mass of black hole is uniform in [3,10] $M_\odot$
\citep{Fryer01,Li15,Cai17}. The beaming angle of $\gamma$-ray bursts are
randomly sampled in interval [0$^\circ$,20$^\circ$]. Since the GW
signal-to-noise ratio in Sec. \ref{sec:SNR} is independent of the
waveform phase, the $\psi$ and $\phi_{(2,0)}$ are not
considered here.

\subsubsection{The signal-to-noise ratio and estimated error}
\label{sec:SNR} A GW signal is claimed to be detected only when
combined signal-to-noise ratio (SNR) $ \ge 8$ for a single detector
network \citep{Sathyaprakash10}. For ET, The combined SNR is
\begin{equation}
    \rho = \sqrt{\sum_{i=1}^{3}(\rho^{(i)})^2},
\end{equation}
where
\begin{equation}
    \rho^{(i)} = \sqrt{\langle\mathcal{H}^{(i)},\mathcal{H}^{(i)} \rangle}.
\end{equation}
and the bracket is defined by
\begin{equation}
    \langle a,b \rangle = 4\int_{f_{\text{min}}}^{f_{\text{max}}}\frac{a(f) b^*(f) + a^*(f) b(f)}{2}\frac{\ud f}{S_h(f)}.
\end{equation}
where $S_h(f)$ is the one-sided noise power spectrum density (PSD), which
determines the performance of a GW detector. We take the noise PSD
of ET to be
\begin{equation}
    S_h(f) = S_0\bigg[x^{p_1} + a_1 x^{p_2} + a_2\frac{1 + b_1 x^1 + b_2 x^2 + b_3 x^3 + b_4 x^4 + b_5 x^5 + b_6 x^6}{1 + c_1 x^1 + c_2 x^2 + c_3 x^3 + c_4 x^4}\bigg],
\end{equation}
as in \cite{Zhao11}, where $x\equiv f/200 $ Hz and $S_0 =
1.449\times 10^{52}$ Hz$^{-1}$. The parameters $p_i$, $a_i$, $b_i$
and $c_i$ are also provided in \cite{Zhao11}. The upper cutoff
frequency $f_{\text{max}}$ is twice the orbit frequency at the last
stable orbit, $f_{\text{max}} = 2f_{\text{LSO}} = 2/(6^{3/2} 2\pi
M_{\text{obs}})$. The lower cutoff frequency is $f_{\text{lower}} =
1$ Hz.

At every simulated redshift, the fiducial value of the luminosity
distance $d_L^{\text{fid}}$ is calculated according to Equation \ref{eqn:1}. Then we
simulate the $\ln d_L^{\text{mea}}$ to be Gaussian distribution
centered around $\ln d_L^{\text{fid}}$ with standard deviation
$\sigma_{\ln d_L}$,
\begin{equation}
\ln d_L^{\text{mea}} = \mathcal{N}(\ln d_L^{\text{fid}},\sigma_{\ln d_L }).
\end{equation}
The fiducial cosmology is flat $\Lambda$CDM cosmology with
$\Omega_{\text{m}} = 0.308, H_0 = 67.8\km \second^{-1}\mpc^{-1}$
\citep{Planck16} when calculating $\ln d_L^{\text{fid}}$.

The Fisher matrix $\Gamma_{ij}$ is widely used to estimate the
errors in the measured parameters,
\begin{equation}
    \Gamma_{ij} = \bigg\langle\frac{\p \mathcal{H}}{\p p_i} ,\frac{\p \mathcal{H}}{\p p_j}\bigg\rangle.
\end{equation}
where $p_i$ denotes the parameters on which the waveforms are
depending, namely $(\ln \mathcal{M}_c, \ln \eta, t_0, \Phi_0,
\cos\iota, \psi, \ln d_L)$. Then the estimated error $\sigma_{p_i}$
of parameter $p_i$ is $(\Gamma^{-1})_{ii}^{1/2}$. However, for
calculation simplicity, we follow \cite{Cai17} and take the distance
uncertainty $\sigma^{\text{inst}}_{d_L}$ to be $2 d_L/\rho$, allowing
for the correlation between $d_L$ and $\iota$. When the additional
error $\sigma^{\text{lens}}_{d_L}$ due to the weak lensing taken
into account, the total uncertainty is
\begin{equation}
\begin{aligned}
    \epsilon_{d_L} &= \sqrt{(\sigma^{\text{inst}}_{d_L})^2 + (\sigma^{\text{lens}}_{d_L})^2}\\
    &= \sqrt{\bigg (\frac{2d_L}{\rho}\bigg )^2 + (0.05z d_L)^2}
\end{aligned}
\end{equation}

\subsubsection{The predicted event rates}
\begin{figure*}
    \centering
    \includegraphics[width=0.7\textwidth]{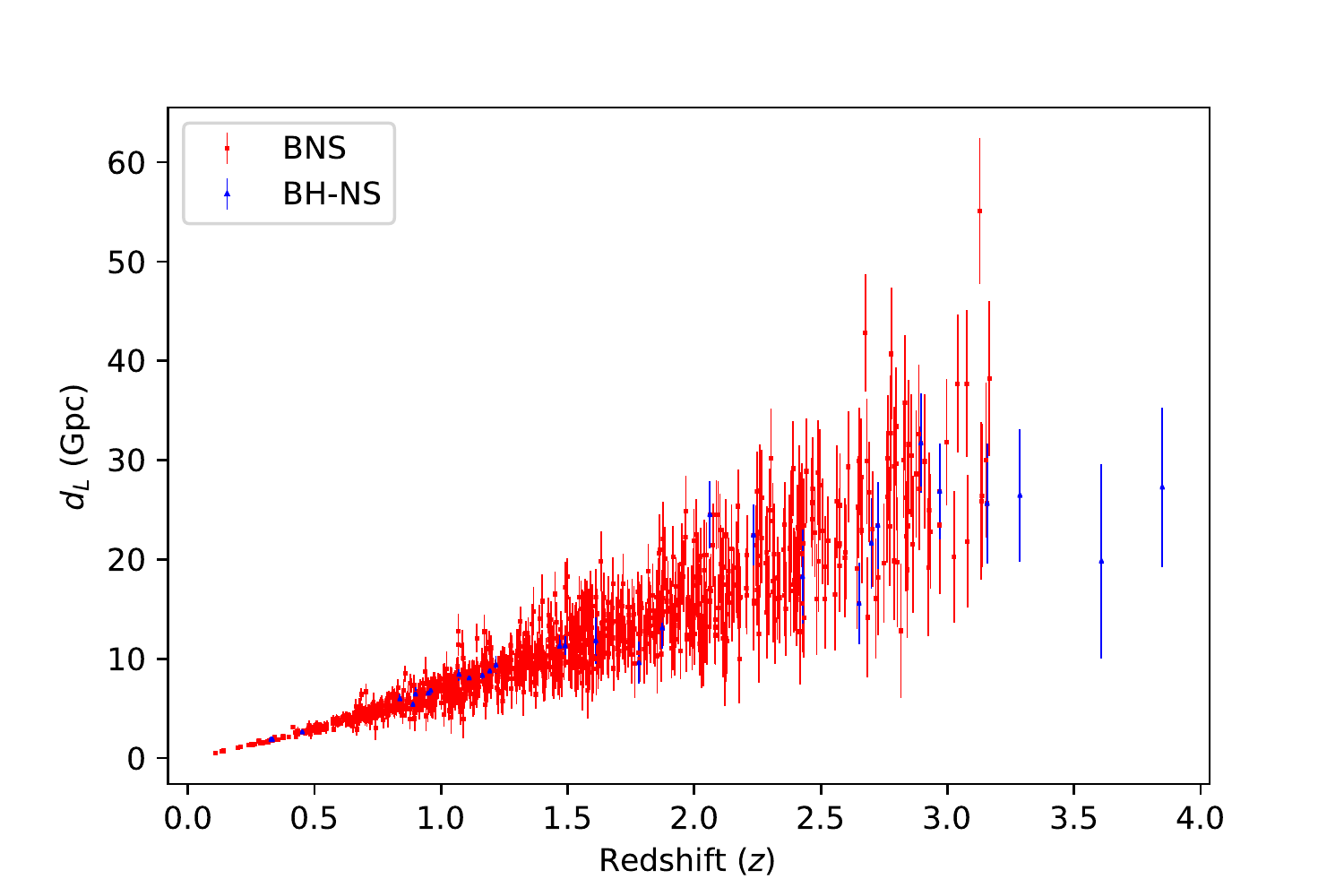}
    \caption{Mock GW catalogue of 1000 BNS merger events and 30 BH-NS merger events as standard sirens.}
    \label{fig:d_L-z}
\end{figure*}

\cite{Abernathy11} predicted event rates in ET. It is expected to
observe $\mathcal{O}(10^3\sim 10^7)$ BNS merger events and
$\mathcal{O}(10^3\sim 10^7)$ BH-NS events per year. However, this
prediction is very uncertain. \cite{Li15} expected that only a small fraction ($\sim 10^{-3}$) of GW detections are accompanied by observed GRBs. Therefore we typically construct a catalogue of
1000 BNS events in our simulation. Besides, when the ratio between NS-BH and BNS events is assumed to be 0.03 as predicted by Advanced LIGO-Virgo network \citep{Abadie10,Li15,Cai17}, 30 NS-BH events are included in the mock catalogue. These simulated events can reach out to a redshift
$z\sim 4$. Figure \ref{fig:d_L-z} shows the $d_L$-$z$ diagram of our
mock GW catalogue.

\section{Calibration of GRB luminosity correlations}

The GRB samples used for calibrating Amati relation
($E_{\text{iso}}$-$E_{\text{p}}$) and Ghirlanda relation
($E_{\gamma}$-$E_{\text{p}}$) are taken from \cite{Wang16} and
\cite{Wang11} respectively.

The energy spectrum of GRBs is modeled by a broken power law
\citep{Band93}
\begin{equation}
\Phi(E) =
\begin{cases}
(\frac{E}{100 \keV})^{\alpha} \exp(-(2 + \alpha) E/E_{\rm p,obs}), & E \le\frac{\alpha -\beta}{2 + \alpha}E_{\rm p,obs}\\
[\frac{(\alpha-\beta)E_{\text{p,obs}}/(2+\alpha)}{100\keV}]^{\alpha-\beta}\exp(\beta-\alpha) (\frac{E}{100\keV})^{\beta}, & {\rm otherwise.}
\end{cases}
\end{equation}
where the typical spectral index values are taken to be $\alpha =
-1.0$ and $\beta = -2.2$ if they are not given in the references.

For each GRB in the sample, the fluence $S$ have been corrected to
1-10000 keV energy band with $k$-correction \citep{Bloom01},
\begin{equation}
    S_{\text{bolo}} = S\times \frac{\int_{1/(1+z)}^{10^4/(1+z)} E\Phi(E)\ud E}{\int_{E_\text{min}}^{E_{\text{max}}} E\Phi(E)\ud E},
\end{equation}
where $E_{\text{min}}$ and $E_{\text{max}}$ are detection thresholds
of the observing instrument. The isotropic energy
$E_{\gamma,\text{iso}}$ and collimation-corrected energy
$E_{\gamma}$ are
\begin{equation}
E_{\gamma,\text{iso}} = \frac{4\pi d_L^2 S_{\text{bolo}}}{1+z},
\end{equation}
and
\begin{equation}
E_{\gamma} = \frac{4\pi d_L^2 S_{\text{bolo}} F_{\text{beam}}}{1+z}.
\end{equation}
respectively, in which $F_{\text{beam}} = 1 -
\cos\theta_{\text{jet}}$ is the beaming factor for jet opening angle
$\theta_{\text{jet}}$. The luminosity distance $d_L$ of low-redshift
GRBs is derived from GW standard sirens using linear interpolation
method \citep{Wang16}, which is independent of cosmology models
\begin{equation}
\ln d_L = \ln d^{\text{GW}}_{L,i} + \frac{z-z_i}{z_{i+1}-z_i}(\ln d^{\text{GW}}_{L,i}-\ln d^{\text{GW}}_{L,i}).
\end{equation}
The $1\sigma$ error can be obtained by
\begin{displaymath}
\sigma_{\ln d_L}^2 = \bigg(\frac{z_{i+1}-z}{z_{i+1}-z_i}\bigg)^2\epsilon_{\ln d^{\text{GW}}_{L, i}}^2 + \bigg(\frac{z-z_i}{z_{i+1}-z_i}\bigg)^2\epsilon_{\ln d^{\text{GW}}_{L, i+1}}^2,
\end{displaymath}
where $\epsilon_{\ln d^{\text{GW}}_L, i}\equiv \epsilon_{d^{\text{GW}}_{L, i}}/ d^{\text{GW}}_{L, i}$ is the distance uncertainty of
the $i$th GW event (the mock GW catalogue has been sorted by
redshift before interpolation).

\subsection{The $E_{\text{iso}}$-$E_{\text{p}}$ correlation}

We parameterize the Amati relation ($E_{\text{iso}}$-$E_{\text{p}}$
correlation) \citep{Amati02} as
\begin{equation}
    \log_{10} \frac{E_{\gamma,\text{iso}}}{1\erg} = a + b\log_{10}\bigg[\frac{E_{\text{p,obs}}(1+z)}{300\keV}\bigg],
\end{equation}
where $E_{\text{p,obs}}(1+z)$ is the cosmological rest-frame
spectral peak energy of GRB. The Markov chain Monte Carlo (MCMC)
algorithm is applied to constrain intercept $a$, slope $b$ and intrinsic scatter $\sigma_{\text{int}}$ of the
correlation. We use the python module $\mathtt{emcee}$ to carry out
parameters fitting \citep{Foreman-Mackey13}. The
likelihood to fit the linear relation $y = ax + b$ \citep{D'Agostini05} is
\begin{equation}
    L = \prod_{i=1}^{N}\frac{1}{\sqrt{2\pi}\sqrt{\sigma_{\text{int}}^2 + \sigma^2_{y_i} + b^2 \sigma^2_{x_i}}} \exp\bigg[\frac{-(y_i - a - b x_i)^2}{2(\sigma_{\text{int}}^2 + \sigma^2_{y_i} + b^2 \sigma^2_{x_i})}\bigg].
\end{equation}
where $x
\equiv \log_{10}[\frac{E_{\text{p,obs}}(1+z)}{300\keV}]$, $y \equiv \log_{10} \frac{E_{\gamma,\text{iso}}}{1\erg}$ and the
propagated uncertainties of $y$ is calculated from
\begin{equation}
\begin{aligned}
    \sigma^2_y = \frac{\sigma^2_{S_{\text{bolo}}}}{(\ln 10\,\, S_{\text{bolo}})^2}.
\end{aligned}
\end{equation}

\subsection{The $E_{\gamma}$-$E_{\text{p}}$ correlation}

The parametrization of the Ghirlanda relation
($E_{\gamma}$-$E_{\text{p}}$ correlation) \citep{Ghirlanda04a} is
\begin{equation}
\log_{10} \frac{E_{\gamma}}{1\erg} = a + b\log_{10}\bigg[\frac{E_{\text{p,obs}}(1+z)}{300\keV}\bigg].
\end{equation}
The likelihood function has the same form as
$E_{\text{iso}}$-$E_{\text{p}}$ correlation's, while the propagated
uncertainties of $y \equiv\log_{10}\frac{E_{\gamma}}{1\erg}$ is
calculated from
\begin{equation}
\begin{aligned}
    \sigma^2_y = \frac{\sigma^2_{S_{\text{bolo}}} F^2_{\text{beam}} + (S_{\text{bolo}})^2\sigma^2_{F_{\text{beam}}}}{(\ln 10\,\, S_{\text{bolo}} F_{\text{beam}})^2}.
\end{aligned}
\end{equation}
The same procedure as handling $E_{\text{iso}}$-$E_{\text{p}}$
correlation is used to calibrate the $E_{\text{iso}}$-$E_{\text{p}}$
correlation.

\subsection{Results}
\begin{figure*}
    \centering
    \includegraphics[width=0.4\textwidth]{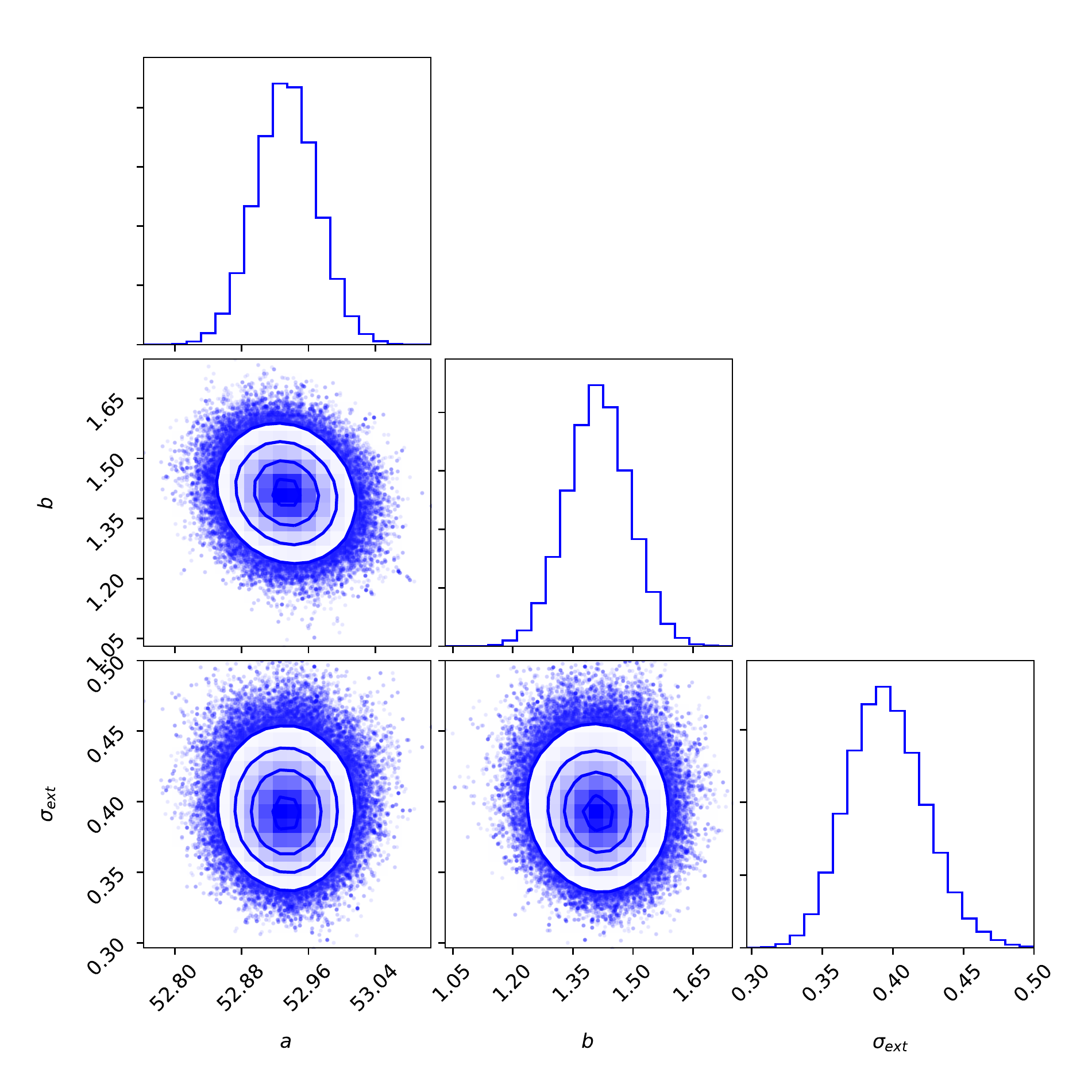}
    \caption{Confidence contours (1$\sigma$, 2$\sigma$ and 3$\sigma$) and marginalized likelihood distributions for intercept $a$ and slope $b$ in Amati relation.}
    \label{fig:Amati}
\end{figure*}

\begin{figure*}
    \centering
    \includegraphics[width=0.4\linewidth]{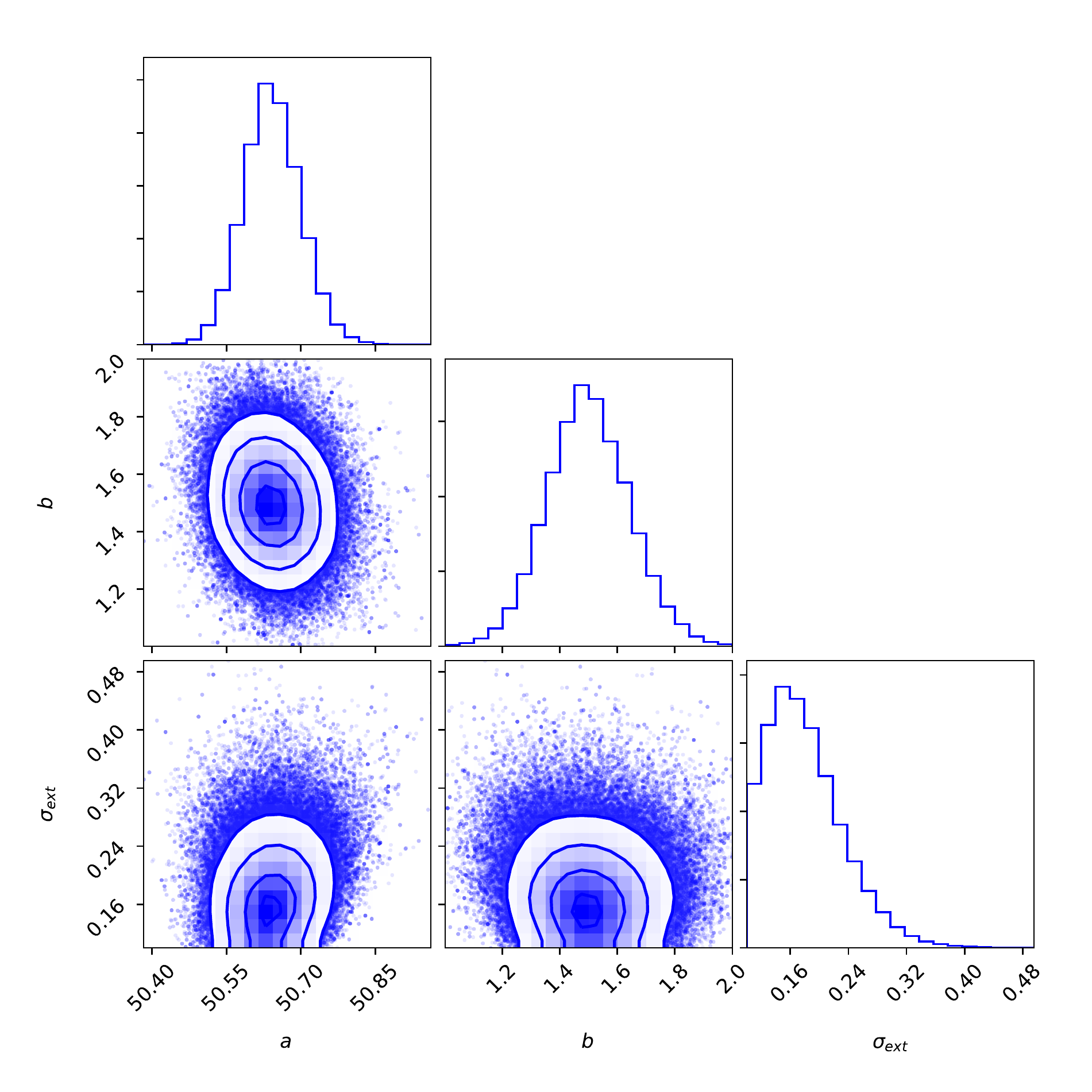}
    \caption{Confidence contours (1$\sigma$, 2$\sigma$ and 3$\sigma$) and marginalized likelihood distributions for intercept $a$ and slope $b$ in Ghirlanda relation.}
    \label{fig:Ghirlanda}
\end{figure*}

With our mock GW catalogue, the constraints on intercept $a$ and
slope $b$ of Amati relation is $a = 52.93\pm 0.04,b = 1.41\pm 0.07$, $\sigma_{\text{int}} = 0.39\pm 0.03$
($1\sigma$) , while for Ghirlanda relation, $a = 50.63\pm 0.08, b =
1.50\pm 0.12$ and $\sigma_{\text{int}} = 0.16\pm 0.04$ ($1\sigma$). The 1$\sigma$, 2$\sigma$ and 3$\sigma$
confidence contours and marginalized likelihood distributions are
shown in Figure \ref{fig:Amati} and Figure \ref{fig:Ghirlanda}
respectively. \cite{Wang16} standardized Amati relation of form $\log_{10}(E_{\gamma,\text{iso}}/\erg) = a + b\log_{10}[E_{\text{p,obs}}(1+z)/\keV]$ with SNe Ia Union2.1 sample. Their fitting results are $a = 48.46\pm 0.033$, $b = 1.766\pm 0.007$ with $\sigma_{\text{ext}} = 0.34\pm 0.04$. \cite{Amati18} calibrated Amati relation of form $\log_{10}(E_{\text{p}}/\keV) = q + m[\log_{10}(E_{\text{iso}}^{\text{cal}}/\erg)-52]$, finding $q = 2.06\pm 0.03$, $m = 0.50\pm 0.12$ and $\sigma_{\text{int}} = 0.20\pm 0.01$. 

\begin{figure*}
	\centering
	\includegraphics[width=0.6\linewidth]{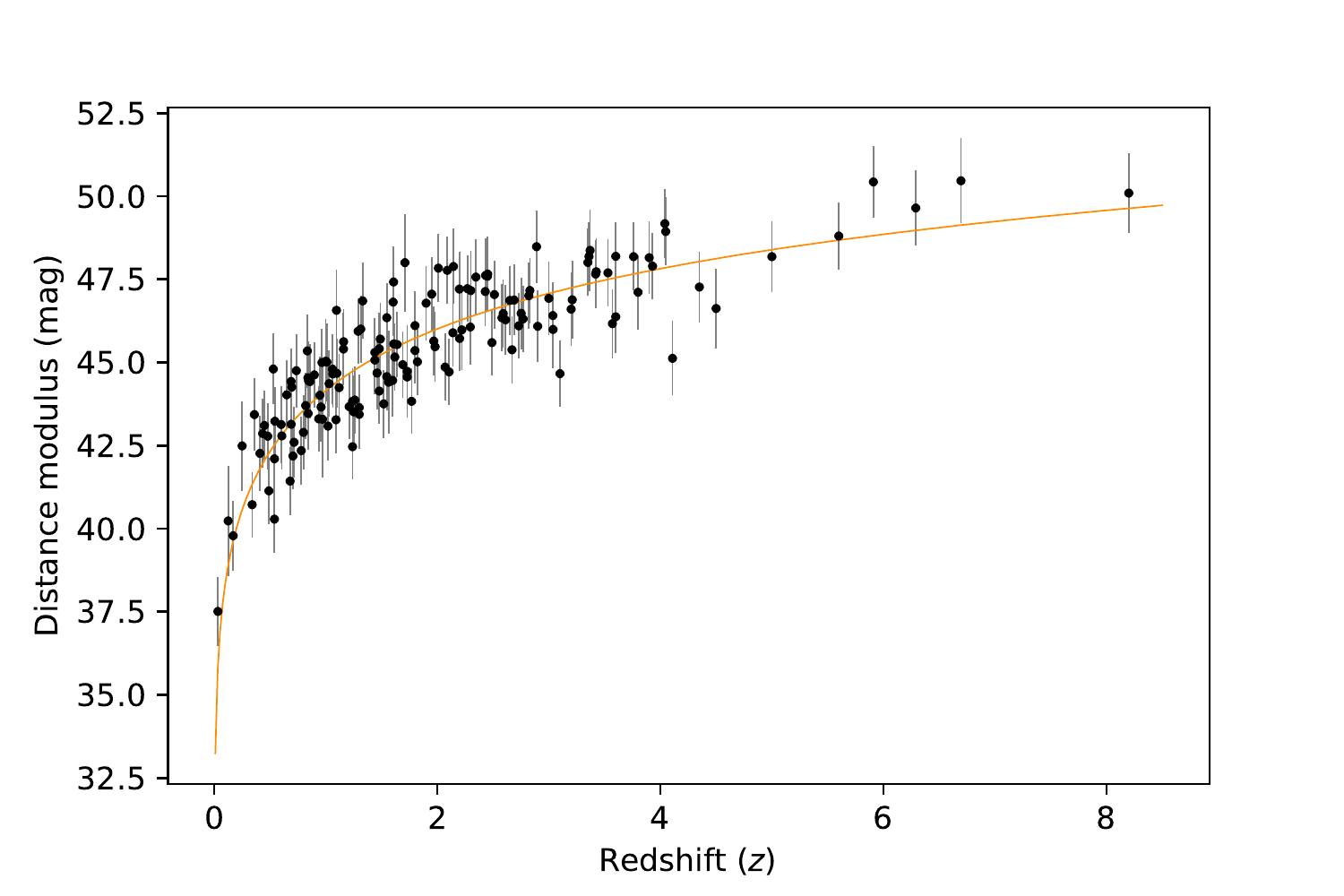}
	\caption{Distance modulus of calibrated GRBs in comparison to Planck15 \citep{Planck16} $\Lambda$CDM cosmology (depicted in solid orange line).}
	\label{fig:Hubble_diagram}
\end{figure*}

The calibrated GRB Hubble diagram is shown in Figure \ref{fig:Hubble_diagram}. The solid line in this figure is plotted based on the Planck15 cosmological parameters \citep{Planck16}.

\subsection{Constraining $\Lambda$CDM model and cosmological applications}
We combine the calibrated GRB data and SNe Ia from Pantheon sample \citep{Scolnic18} to constrain non-flat $\Lambda$CDM model. The nuisance parameters \{$\alpha$, $\beta$, $M_B^1$, $\Delta_M$\} of SNe Ia lightcurve are fitted with cosmological parameters \{$\Omega_{\text{m}}$, $\Omega_{\Lambda}$\} simultaneously with the following total likelihood function
\begin{displaymath}
L \propto L_{\text{GRB}}\cdot L_{\text{SN}},
\end{displaymath}
The likelihood function of GRB is given by 
\begin{equation}
	L_{\text{GRB}} = \prod_{i=1}^{N}\frac{1}{\sqrt{2\pi}\sigma_{\mu_{\text{GRB},i}}} \exp\bigg[\frac{-(\mu_{\text{th},i} - \mu_{\text{GRB},i})^2}{2\sigma_{\mu_{\text{GRB},i}}^2}\bigg].
\end{equation}
where the distance modulus uncertainty $\sigma_{\mu_{\text{GRB}}}$ is
\begin{equation}
\sigma_{\mu_{\text{GRB}}}^2 = \bigg(\frac{5}{2}\sigma_{\log_{10} E_{\gamma,\text{iso}}}\bigg)^2 + \bigg(\frac{5}{2\ln 10}\frac{\sigma_{S_{\text{bolo}}}}{S_{\text{bolo}}}\bigg)^2
\end{equation}
and
\begin{equation}
\sigma_{\log_{10} E_{\gamma,\text{iso}}}^2 = \sigma_a^2 + \bigg(\sigma_b\log_{10}\frac{E_\text{p,obs}(1+z)}{300\keV}\bigg)^2 + \bigg(\frac{b}{\ln 10}\frac{\sigma_{E_{\text{p,obs}}}}{E_{\text{p,obs}}}\bigg)^2 + \sigma_{\text{int}}^2
\end{equation}

\begin{figure*}
	\centering
	\includegraphics[width=0.6\linewidth]{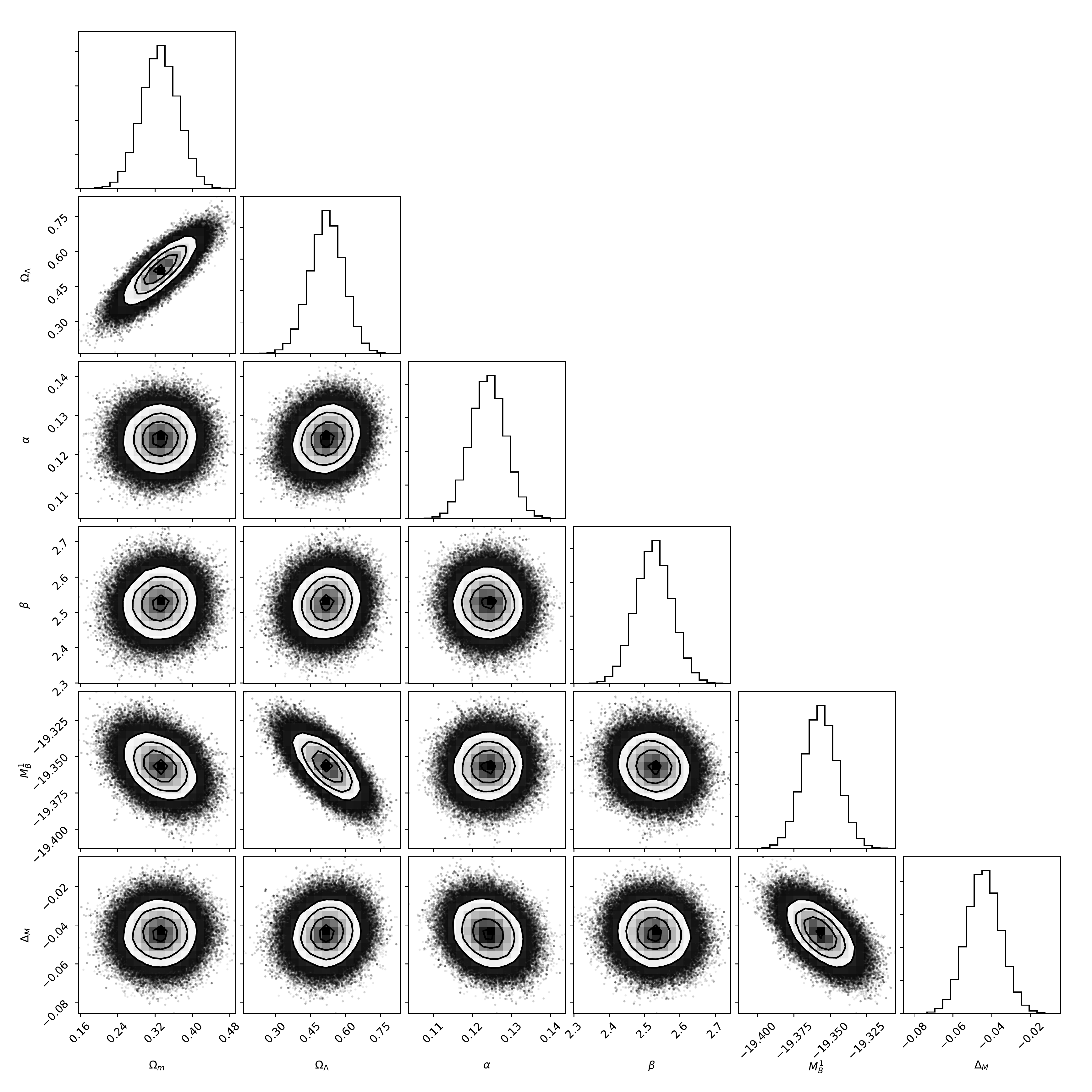}
	\caption{1$\sigma$, 2$\sigma$ and 3$\sigma$ constraints on $\Omega_{\text{m}}$, $\Omega_{\Lambda}$ and SNe Ia lightcurve parameters \{$\alpha$, $\beta$, $M_B^1$, $\Delta_M$\} from "calibrated" GRB and Pantheon SNe Ia sample.}
	\label{fig:LCDM}
\end{figure*}

The Hubble constant $H_0$ in our fitting is fixed to Planck15 \citep{Planck16} value. With the combined sample (GRBs + SNe), the best-fit values for non-flat $\Lambda$CDM model are $\Omega_{\text{m}} = 0.33\pm 0.04$ and $\Omega_{\Lambda} = 0.52\pm 0.08$ with 1$\sigma$ uncertainties. The constraints on $\Omega_{\text{m}}$, $\Omega_{\Lambda}$ and SNe Ia lightcurve parameters \{$\alpha$, $\beta$, $M_B^1$, $\Delta_M$\} are shown in Figure \ref{fig:LCDM}.

\section{Summary}
In this paper, we propose to calibrate the GRB luminosity relations
using GW standard sirens. This method is model-independent and will
overcome the circularity problem. The constraints for intercepts and
slopes of Amati relation and Ghirlanda relation are $a = 52.93\pm
0.04,b = 1.41\pm 0.07$, $\sigma_{\text{int}} = 0.39\pm 0.03$ ($1\sigma$) and $a = 50.63\pm 0.08, b =
1.50\pm 0.12$, $\sigma_{\text{int}} = 0.16\pm 0.04$ ($1\sigma$) respectively with our mock GW catalogue.
The performance of our method will improve with the upgrade of GW
detector's sensitivity, especially with third generation detectors
ET \citep{Abernathy11}, BBO \citep{Cutler09} and DECIGO
\citep{Kawamura11}.
GRBs serve as a complementary tool to other cosmological probes
such as SNe Ia, BAO and CMB. Besides, it
plays a crucial role in constraining $w(z)$ especially at high
redshifts \citep{Wang11}, which may help us understanding the nature
of dark energy.

\acknowledgments
We thank the anonymous referee for constructive comments. We thank Wen Zhao, Tao Yang and Jun-Jie Wei for helpful suggestions. This work is supported by the National Natural Science Foundation of China (grant U1831207).

\end{document}